\documentclass[conference]{IEEEtran}

\usepackage{amsmath,amssymb}
\usepackage{hyperref}
\usepackage{okcategory,okgeneralmath,oktheorem}
\usepackage{stmaryrd}
\usepackage{tikz}

\usetikzlibrary{cd}

\title{A monadic solution to the Cartwright-Felleisen-Wadler conjecture}

\author{\IEEEauthorblockN{Ohad Kammar}
\IEEEauthorblockA{University of Oxford\\
\url{Ohad.Kammar@cs.ox.ac.uk}}
\and
\IEEEauthorblockN{Dylan McDermott (speaker)\\
University of Cambridge}
\IEEEauthorblockA{\url{Dylan.McDermott@cl.cam.ac.uk}}}

\newcommand\C{\mathcal{C}}
\newcommand\E{\mathcal{E}}
\newcommand\M{\mathcal{M}}
\newcommand\op{\mathrm{op}}

\begin{document}

\maketitle{}

Given a programming language, can we give a monadic semantics that is stable under language extension? This problem was posed independently in
Cartwright and
Felleisen~\cite{cartwright-felleisen:extensible-denotational-language-specifications}
and
Wadler~\cite{wadler:marriage-effects-monads}.
It is yet to be resolved. In this talk, we describe a solution to it
based on factorizations of monad morphisms.

\section{The conjecture}
\label{section:conjecture}
First we will give a more detailed description of the problem. Consider
a programming language with effects. A model for this language consists
of a category $\C$ and a monad $T$ over $\C$, and each operation $\op
: A \rightarrow B$ has a corresponding Kleisli arrow
$\llbracket \op\rrbracket : \llbracket A \rrbracket \rightarrow
T\,\llbracket B \rrbracket$. A simple example is global state with the
monad $T = S \rightarrow - \times S$ and operations $\mathsf{read} :
\mathsf{unit} \rightarrow S$ and $\mathsf{write} : \mathsf{S} \rightarrow
\mathsf{unit}$. In general, the monad $T$ may need to combine many
different effects, such as nondeterminism, I/O, and local state.

Models such as these are
\emph{unstable} under the addition of language features: if we decide to
add a new effect to the language, such as exceptions, we may need a new
monad, even to model the parts of the language that do not use the new
effect. All expressions will be given different
denotations than previously. Even expressions that do not use
the new effects receive more complicated denotations. For example, if we
started with a read-only state and then add the ability to write, we
switch from the reader monad $S \rightarrow -$ to the global state
monad. This is undesirable. To use a slogan: simpler programs should retain their
simpler semantics.

Cartwright and
Felleisen~\cite{cartwright-felleisen:extensible-denotational-language-specifications}
noticed this problem. The (non-monadic) semantics they
describe addresses it.

We can rephrase the problem using \emph{effect systems}, in which
simpler fragments are
restricted to a subsignature $\epsilon$ of operations.
Wadler~\cite{wadler:marriage-effects-monads} proposed using monads $T_\epsilon$ over $\C$ and a Kleisli arrow
$\llbracket\op\rrbracket_\epsilon : \llbracket A \rrbracket \rightarrow
T_\epsilon\,\llbracket B \rrbracket$ for each $\op\in\epsilon$. A model
would have one monad per set $\epsilon$. Now if
we add new effects to the language, expressions that only use the
original effects will be given the same denotations, since we can
interpret them using the original monad. Models for parts of the
language can be simpler than the model for the entire language, because
they do not need to support all of the operations in the language. If the
language has global state, the part that can only write (corresponding to
the subsignature $\epsilon = \{\mathsf{write}\}$) can be modelled using
the monad $(\mathbf{1} + S) \times -$, and this will still be true even if
we add other effects to the language. Choosing the correct $T_\epsilon$
also gives us an easier way to reason about the part of the language corresponding
to $\epsilon$. Wadler~\cite{wadler:marriage-effects-monads} conjectures that there should be some method for
constructing models of this form using $T$.

The problem that we consider in this talk is whether we can construct
models of the second kind from models of the first.
Given the monad $T$ we want, for every subsignature
$\epsilon$, a monad $T_\epsilon$ that allows us to
interpret the operations in $\epsilon$. The resulting model would
automatically be stable and should
satisfy two additional properties:
\begin{itemize}
  \item{Correctness: it should agree with $T$ in the sense that it
    should identify the same programs as the original model.}
  \item{Modularity: each $T_\epsilon$ should support only the
    operations in $\epsilon$. In other words, the monads should not
    be more complicated than necessary.}
\end{itemize}
The construction should also be as general as possible.
By solving this problem, we answer the conjecture made by Wadler, and address the problems with monadic semantics discussed by
Cartwright and Felleisen. The solution is useful because it allows us
to, for example, validate effect-dependent program
transformations~\cite{kammar-plotkin:algebraic-foundations-effect-optimisations,benton-kennedy-beringer-hofmann:relational-semantics-program-transformations}.
This is because we can reason about the simpler monads $T_\epsilon$
instead of $T$.

There have already been partial solutions to this problem.
Katsumata~\cite{katsumata:graded-monads} describes a method of
constructing \emph{graded monads}. When the grading is by sets of
operations and the monad is free the construction creates a monad for each subsignature.
It is not clear that the same construction can be used for non-free
monads, and hence it is not as general as we would like.

Kammar's thesis~\cite{kammar:thesis} describes a construction that does not
require any additional data, based on factorizations of morphisms of
Lawvere theories. Apart from its increased mathematical sophistication,
it does not apply to all of the cases we would like to
consider, since it requires the semantics to be algebraic.
Continuations, for example, are excluded. The construction we describe
generalizes it by considering factorizations of monad morphisms.

\section{Constructing stable models}
To describe the construction, we use the standard notion of
\emph{factorization system}~\cite{adamek-herrlich-strecker:joy-of-cats}.
A factorization system $\pair{\E}{\M}$ for a category $\C$ consists of
two classes of morphisms $\E$ and $\M$ such that each morphism $f : X
\rightarrow Y$ can be factored into $f = n \compose e$, for some
$n\in\M$ and $e\in\E$, and some additional properties are satisfied.

There are many examples of factorization systems. For $\Set$, we can
take $\E$ as the class of surjective functions and $\M$ as the class of
injective functions. Every function factorizes into a surjection
followed by an inclusion. For presheaves we take pointwise surjections
and injections.  For $\wCPO$ we take $\E$ as the class of dense
epimorphisms (continuous functions where the codomain is the closure of
the image) and $\M$ as the class of full monomorphisms (continuous
functions $f$ satisfying $f\,x \leq f\,y \Rightarrow x \leq y$). We will
assume from now on that the category $\C$ has a given factorization
system $\pair{\E}{\M}$.

The main theorem allows us to factor monad morphisms. It observes that if we
have a monad $S$ and $\E$ is closed under products and $S$ (i.e. if
$e\in\E$ then $e \times \mathrm{id}\in\E$ and $S\,e\in\E$), then monad
morphisms $S \rightarrow T$ can be factored pointwise. We do not make any additional assumptions about $T$.

\begin{theorem}
\label{factorization-theorem}
Let $m : S \rightarrow T$ be a strong monad morphism, and factorize $m$ pointwise:
\[
\begin{tikzcd}
  S\,X
  \arrow[rr, "m_X"]
  \arrow[rd, "e_X" below left, two heads] & & T X \\
  &
  R\,X
  \arrow[ru, "n_X" below right, tail]
\end{tikzcd}
\]
If $\E$ is closed under $S$ and products then $R$ is a strong monad $e$
and $n$ are strong monad morphisms.
\end{theorem}

The intuition behind the monad $R$ is that computations do not satisfy
more equations than $T$ since $n$ is in $\M$ (and elements of $\M$ act
like injections). They also do not support
more operations than $S$ because $e$ is in $\E$ (and elements of $\E$
act like surjections). Many monads $S$ satisfy the precondition given in
the theorem, and therefore this provides a general method of
constructing simpler monads from more complex ones.

We are particularly interested in factorizations yielding the simplest
possible $T_\epsilon$. We
choose $S$ to be the free
monad with operations in $\epsilon$ and $T_\epsilon$ to be the induced $R$. It behaves the same as $T$ (in
the sense that it satisfies the same equations), and it is not more
complex than necessary (the only operations it supports are those in
$\epsilon$).
The Kleisli arrows
$\llbracket \mathsf{op} \rrbracket_\epsilon$ are given by the
composition of $e$ and the interpretation of $\mathsf{op}$ using the
free monad. The following lemma allows us to use the free monad when it
is constructed as a colimit (see
\cite{kelly:transfinite-constructions}).

\begin{lemma}
\label{free-monad-lemma}
Let $\kappa$ be a regular cardinal and $F : \C \rightarrow \C$ be an
endofunctor. If $\C$ has $\kappa$-directed colimits and $F$ preserves
them then $\E$ is closed under the free monad for $F$.
\end{lemma}

The free monad for the subsignature $\epsilon$ is given by taking $F$ to be
\[F = \sum_{(\mathrm{op} : A \rightarrow B) \in \epsilon} A \times (-)^B \]
This functor often preserves $\kappa$-directed colimits in practice.

In summary, the construction works as follows. We assume that we have
a factorization system for which $\E$ is closed under products. First we construct the
functor $F$ from the subsignature $\epsilon$. $F$ should preserve $\kappa$-directed
colimits (so that it can be used in Lemma~\ref{free-monad-lemma}). We then construct the free
monad $S$ for the functor.  Finally, we
factorize the unique monad morphism $m : S \rightarrow T$ as in
Theorem~\ref{factorization-theorem} to get
$T_\epsilon$. We can interpret each of the operations in $\epsilon$
using $T_\epsilon$ as described above.

In ongoing work we apply the construction to languages with a
range of different effects, such as state, names, probability and
continuations. We also plan to show that this construction does indeed
generalize the construction given by Kammar~\cite{kammar:thesis}. The
construction is sufficiently general to apply to a wide range of
languages.

\section{Correctness}
We have described a method of constructing simpler monads from more
complex ones. We have yet to show that the results of this construction
satisfy the correctness and modularity properties from
Section~\ref{section:conjecture}. We now briefly describe some
preliminary work for showing that the construction is
correct.

We need to show that $T_\epsilon$ behaves in the same way as $T$. To
do this we will use a logical relations proof.
Katsumata~\cite{katsumata:relating-computational-effects} introduces a
notion of \emph{fibration for logical relations}. This provides a notion
of predicate that can be used to construct suitable logical relations.

Factorization systems also provide a notion of predicate for a category:
a morphism $X \rightarrowtail Y$ in $\M$ can be seen as a predicate that
is defined on $Y$ and true on $X$. Since we already assume that the
category has a factorization system, we would like to use it as the
fibration for logical relations.

Hughes and Jacobs~\cite{hughes-jacobs:factorization-systems-fibrations}
determine the precise relationship between factorization systems and
fibrations. We extend this correspondence to determine the relationship between
factorization systems and fibrations for logical relations. In
particular, we show that factorization systems with certain additional
properties induce fibrations for logical relations. These additional
properties are satisfied by all of the factorization systems we are
interested in. Hence we get a way to prove the correctness of the
construction for free from the factorization system.

\bibliographystyle{plain}
\bibliography{proposal}

\end{document}